\documentclass[a4paper,11pt]{article}

\usepackage{hyperref}%
\hypersetup{colorlinks=true,%
  linkcolor=blue,%
  citecolor=magenta,%
  urlcolor=black,%
  bookmarksnumbered=true,%
  bookmarkstype=toc,%
  bookmarksopen=false,%
  pdftitle={Potential Algebra Approach to Quantum Mechanics with
    Generalized Uncertainty Principle},%
  pdfkeywords={minimal length uncertainty; shape invariance; potential
    algebra},%
  pdfauthor={Satoshi Ohya and Pinaki Roy}}

\usepackage[margin=1in]{geometry}

\usepackage[bitstream-charter,cal=cmcal]{mathdesign}
\usepackage[T1]{fontenc}

\usepackage{xcolor}

\usepackage{amsmath,amsbsy}

\usepackage{epic,eepic}

\usepackage[hang,font=small,labelfont=bf]{caption}%

\makeatletter%
\@addtoreset{equation}{section}%
\makeatother%

\usepackage{titlesec}
\titleformat{\section}[block]{\filright\bfseries}{\thesection.}{0.5em}{}[]
\titleformat{\subsection}[block]{\filright\bfseries}{\thesubsection.}{0.5em}{}

\usepackage{cite}

\title{\large\bfseries Potential Algebra Approach to Quantum
  Mechanics\\ with Generalized Uncertainty Principle}
\author{
  {\normalsize Satoshi Ohya${}^{1,}$\footnote{\texttt{ohya@phys.cst.nihon-u.ac.jp}}~~~and~~~Pinaki Roy${}^{2,3,}$\footnote{\texttt{pinaki.roy@tdtu.edu.vn}}}\\[1em]
  {\small\itshape ${}^{1}$Institute of Quantum Science, Nihon University}\\
  {\small\itshape Kanda-Surugadai 1-8-14, Chiyoda, Tokyo 101-8308, Japan}\\
  {\small\itshape ${}^{2}$Atomic~Molecular~and~Optical~Physics~Research~Group,}\\ {\small\itshape Advanced Institute of Materials Science, Ton Duc Thang University, Ho~Chi~Minh~City, Vietnam}\\
  {\small\itshape ${}^{3}$ Faculty of Applied Sciences, Ton Duc Thang
    University, Ho~Chi~Minh~City,~Vietnam}}
\date{\small (Dated: \today)}

\begin{document}
\maketitle
\flushbottom

% ABSTRACT %
\begin{abstract}
  In this note, we study the potential algebra for several models
  arising out of quantum mechanics with generalized uncertainty
  principle. We first show that the eigenvalue equation corresponding
  to the momentum-space Hamiltonian
  \[H=-(1+\beta p^{2})\frac{d}{dp}(1+\beta
    p^{2})\frac{d}{dp}+g(g-1)\beta^{2}p^{2}-g\beta,\] which is
  associated with some one-dimensional models with minimal length
  uncertainty, can be solved by the unitary representations of the Lie
  algebra $\mathfrak{su}(2)$ if
  $g\in\{\tfrac{1}{2},1,\tfrac{3}{2},2,\cdots\}$. We then apply this
  result to spectral problems for the non-relativistic harmonic
  oscillator as well as the relativistic Dirac oscillator in the
  presence of a minimal length and show that these problems can be
  solved solely in terms of $\mathfrak{su}(2)$.
\end{abstract}

\newpage
\section{Introduction}
\label{section:1}
Study of symmetry structure of quantum mechanical models is
interesting as well as useful as one may determine various observables
using symmetry properties \cite{Bohm:1988hq}. In this respect,
potential algebra is a powerful tool to obtain the spectrum and
scattering amplitude of quantum mechanical models in a purely
algebraic fashion \cite{Alhassid:1983toa}. On the other hand, in
supersymmetric quantum mechanics, which is closely related to the
factorization method \cite{Infeld:1951mw}, the concept of shape
invariance \cite{Gendenshtein:1984vs} plays a very important role in
obtaining solutions without solving differential equations:
shape-invariant potentials allow complete determination of the
spectrum and eigenfunctions in a purely algebraic fashion. In fact, it
has been discussed that these two approaches are essentially
equivalent \cite{Gangopadhyaya:1998ccj} (see also
\cite{Balantekin:1997mg}).

The purpose of this note is to examine energy spectrum of some quantum
mechanical models arising out of the minimal length uncertainty
formalism \cite{Kempf:1994su} from the viewpoint of potential
algebra. Here the minimal length uncertainty formalism is the simplest
method to incorporate the fundamental length scale (such as the Planck
length) into the realm of quantum mechanics. Since it just modifies
the Heisenberg uncertainty relation, the minimal length uncertainty
formalism has been attracted much attention in order to understand the
effect of the presence of the fundamental length scale without
invoking quantum gravity. In the rest of this note we shall
investigate eigenvalue problems for Hamiltonians of some minimal
length models by using the potential algebra technique. To be more
specific, in this note we shall focus on the following two-parameter
family of momentum-space Hamiltonian:
\begin{align}
  H=-(1+\beta p^{2})\frac{d}{dp}(1+\beta p^{2})\frac{d}{dp}+g(g-1)\beta^{2}p^{2}-g\beta,\label{eq:1.1}
\end{align}
where $\beta$ and $g$ are parameters of the model. As discussed in
\cite{Spector:2007}, this Hamiltonian is known to be shape
invariant. We shall show that the potential algebra associated with
\eqref{eq:1.1} is nothing but the Lie algebra $\mathfrak{su}(2)$ and
determine the spectrum only through the unitary representations of
$\mathfrak{su}(2)$. Subsequently, we shall consider a number of
minimal length uncertainty models and identify them with the
Hamiltonian \eqref{eq:1.1} by choosing the parameter $g$ suitably. As
illustrative examples, we shall focus on the non-relativistic harmonic
oscillator and the relativistic Dirac oscillator subject to the
minimal length uncertainty principle. The Hamiltonians of these models
are respectively given by
\begin{subequations}
  \begin{align}
    H&=\frac{1}{2m}p^{2}+\frac{1}{2}m\omega^{2}x^{2},\label{eq:1.2a}\\
    H&=c\sigma_{y}(p-i\sigma_{z}m\omega x)+\sigma_{z}mc^{2},\label{eq:1.2b}
  \end{align}
\end{subequations}
whose energy eigenvalues are known to be of the following forms
\cite{Kempf:1994su,Nouicer:2006xua}:
\begin{subequations}
  \begin{align}
    E_{n}&=\hbar\omega\left(n+\frac{1}{2}\right)\left[\frac{m\hbar\omega\beta}{2}+\sqrt{1+\left(\frac{m\hbar\omega\beta}{2}\right)^{2}}\right]+\frac{1}{2}m\hbar^{2}\omega^{2}\beta n^{2},\label{eq:1.3a}\\
    E_{n}&=\pm mc^{2}\sqrt{1+\frac{\hbar^{2}\omega^{2}\beta}{c^{2}}n^{2}+\frac{2\hbar\omega}{mc^{2}}n},\label{eq:1.3b}
  \end{align}
\end{subequations}
where $n\in\{0,1,2,\cdots\}$ and $\hbar\sqrt{\beta}$ is the minimal
length uncertainty. The goal of this note is to show that the energy
spectrum \eqref{eq:1.3a} and \eqref{eq:1.3b}, though look quite
different, can be completely determined through the unitary
representations of the Lie algebra $\mathfrak{su}(2)$. Before going
into this, however, let us first briefly recall the basics of the
minimal length uncertainty principle and the shape invariance of the
Hamiltonian \eqref{eq:1.1}.

\section{Minimal length uncertainty principle and shape invariance}
\label{section:2}
To begin with, let us first recall that position ($x$) and momentum
($p$) of particles whose length cannot be measured below a minimum
value satisfy the following commutation relation \cite{Kempf:1994su}:
\begin{equation}
  [x,p]=i\hbar(1+\beta p^{2}),\label{eq:2.1}
\end{equation}
where $\hbar\sqrt{\beta}$ ($\beta>0$) gives the minimal length
uncertainty. The corresponding uncertainty relation reads
\begin{equation}
  \Delta x\Delta p\geq\frac{\hbar}{2}\left[1+\beta(\Delta p)^{2}+\beta\langle p\rangle^{2}\right].\label{eq:2.2}
\end{equation}
A realization of the operators $x$ and $p$ which satisfy the
commutation relation \eqref{eq:2.1} can be taken as
\begin{equation}
  x=i\hbar(1+\beta p^{2})\frac{d}{dp}\quad\text{and}\quad p=p.\label{eq:2.3}
\end{equation}
One of the most important features of this class of models is that the
inner product should be given by
\begin{equation}
  \langle\psi|\phi\rangle=\int_{-\infty}^{\infty}\frac{dp}{1+\beta p^{2}}\,\psi^{\ast}(p)\phi(p),\label{eq:2.4}
\end{equation}
under which the operators $x$ and $p$ in \eqref{eq:2.3} become
(formally) hermitian.

Now, following ref.~\cite{Spector:2007} let us consider the following
first-order differential operators in momentum
space:\footnote{Eqs.~\eqref{eq:2.5a} and \eqref{eq:2.5b} correspond to
  the choice $a=1$, $b=\beta$, and $c=g\beta$ in \cite{Spector:2007}.}
\begin{subequations}
  \begin{align}
    A(g)&=+(1+\beta p^{2})\frac{d}{dp}+g\beta p=+(1+\beta p^{2})^{1-\frac{g}{2}}\frac{d}{dp}(1+\beta p^{2})^{+\frac{g}{2}},\label{eq:2.5a}\\
    \Bar{A}(g)&=-(1+\beta p^{2})\frac{d}{dp}+g\beta p=-(1+\beta p^{2})^{1+\frac{g}{2}}\frac{d}{dp}(1+\beta p^{2})^{-\frac{g}{2}},\label{eq:2.5b}
  \end{align}
\end{subequations}
where $g$ is a dimensionless real parameter. Notice that these
operators are (formally) hermitian conjugate with each other with
respect to the inner product \eqref{eq:2.4}. Let us next introduce the
following factorized Hamiltonians in momentum space:
\begin{subequations}
  \begin{align}
    H(g)&=\Bar{A}(g)A(g)=-(1+\beta p^{2})\frac{d}{dp}(1+\beta p^{2})\frac{d}{dp}+g(g-1)\beta^{2}p^{2}-g\beta,\label{eq:2.6a}\\
    \Tilde{H}(g)&=A(g)\Bar{A}(g)=-(1+\beta p^{2})\frac{d}{dp}(1+\beta p^{2})\frac{d}{dp}+g(g+1)\beta^{2}p^{2}+g\beta,\label{eq:2.6b}
  \end{align}
\end{subequations}
both of which are (formally) hermitian with respect to the inner
product \eqref{eq:2.4}. It should be noted that these Hamiltonians
satisfy the following identity (translational shape invariance):
\begin{align}
  \Tilde{H}(g)=H(g+1)+[2(g+1)-1]\beta.\label{eq:2.7}
\end{align}

Now we wish to solve the following eigenvalue equation for the
Hamiltonian $H(g)$:
\begin{align}
  H(g)\psi(p)=E\psi(p).\label{eq:2.8}
\end{align}
By using the relation \eqref{eq:2.7} the energy eigenvalues are
readily found to be of the form \cite{Spector:2007}
\begin{align}
  E_{n}=\sum_{k=1}^{n}[2(g+k)-1]\beta=(n^{2}+2ng)\beta,\quad n\in\{0,1,2,\cdots\}.\label{eq:2.9}
\end{align}
In what follows we shall show that the eigenvalue problem
\eqref{eq:2.8} can be solved solely in terms of the Lie algebra
$\mathfrak{su}(2)$.

\section{Potential algebra}
\label{section:3}
The purpose of this note is to understand the (Lie-)algebraic
structure behind the spectral problem for the Hamiltonian
\eqref{eq:1.1}. To this end, we would like to translate the shape
invariance \eqref{eq:2.7} into the language of potential algebra. To
the best of our knowledge, there have been proposed two seemingly
different approaches to the (Lie-)algebraic description of shape
invariance. The first is due to Balantekin \cite{Balantekin:1997mg},
and the second is due to Gangopadhyaya \textit{et al.}
\cite{Gangopadhyaya:1998ccj}. As briefly discussed in appendix
\ref{appendix:A}, however, these two approaches are essentially
equivalent and give the same result in the present problem. In this
note, we will follow (with slight modifications) the prescription
given in \cite{Gangopadhyaya:1998ccj} and introduce the potential
algebra as follows. We first introduce an auxiliary periodic variable
$\theta\in[0,2\pi)$ and consider the following first-order
differential operators:
\begin{subequations}
  \begin{align}
    J_{z}&=-i\partial_{\theta},\label{eq:3.1a}\\
    J_{+}&=\mathrm{e}^{+i\theta}A(J_{z})=\mathrm{e}^{+i\theta}\left[+(1+\beta p^{2})\partial_{p}+\beta pJ_{z}\right],\label{eq:3.1b}\\
    J_{-}&=\Bar{A}(J_{z})\mathrm{e}^{-i\theta}=\left[-(1+\beta p^{2})\partial_{p}+\beta pJ_{z}\right]\mathrm{e}^{-i\theta}.\label{eq:3.1c}
  \end{align}
\end{subequations}
Note that $J_{z}$ is (formally) hermitian and $J_{\pm}$ are (formally)
hermitian conjugate with each other with respect to the inner product
\begin{align}
  \langle\psi|\phi\rangle=\int_{-\infty}^{\infty}\!\frac{dp}{1+\beta p^{2}}\int_{0}^{2\pi}\!\!d\theta\,\psi^{\ast}(p,\theta)\phi(p,\theta).\label{eq:3.2}
\end{align}
Let us next study the commutation relations of the operators
\eqref{eq:3.1a}--\eqref{eq:3.1c}. By using the identities
$J_{+}=\mathrm{e}^{i\theta}A(J_{z})=A(J_{z}-1)\mathrm{e}^{i\theta}$
and
$J_{-}=\Bar{A}(J_{z})\mathrm{e}^{-i\theta}=\mathrm{e}^{-i\theta}\Bar{A}(J_{z}-1)$,
which follow from the relation
$\mathrm{e}^{i\theta}J_{z}\mathrm{e}^{-i\theta}=J_{z}-1$, we get
\begin{align}
  J_{+}J_{-}
  &=A(J_{z}-1)\Bar{A}(J_{z}-1)=\Tilde{H}(J_{z}-1)=H(J_{z})+[2(J_{z}-1)+1]\beta\nonumber\\
  &=\Bar{A}(J_{z})A(J_{z})+2\beta\left(J_{z}-\frac{1}{2}\right)=J_{-}J_{+}+2\beta\left(J_{z}-\frac{1}{2}\right),\label{eq:3.3}
\end{align}
where we have used the relations $H(J_{z})=\Bar{A}(J_{z})A(J_{z})$ and
$\Tilde{H}(J_{z})=A(J_{z})\Bar{A}(J_{z})$. The third equality follows
from the shape invariance condition \eqref{eq:2.7}. Similarly, a
straightforward calculation gives
\begin{align}
  J_{z}J_{\pm}=J_{\pm}J_{z}\pm J_{\pm}.\label{eq:3.4}
\end{align}
We thus find the following commutation relations:
\begin{subequations}
  \begin{align}
    [J_{+},J_{-}]&=2\beta\left(J_{z}-\frac{1}{2}\right),\label{eq:3.5a}\\
    [J_{z},J_{\pm}]&=\pm J_{\pm}.\label{eq:3.5b}
  \end{align}
\end{subequations}
Note that these commutation relations are essentially equivalent to
those of the Lie algebra $\mathfrak{su}(2)$. Indeed, by redefining the
operators as
\begin{subequations}
  \begin{align}
    \Tilde{J}_{z}&=J_{z}-\frac{1}{2},\label{eq:3.6a}\\
    \Tilde{J}_{\pm}&=\frac{1}{\sqrt{\beta}}J_{\pm},\label{eq:3.6b}
  \end{align}
\end{subequations}
one immediately finds that \eqref{eq:3.6a} and \eqref{eq:3.6b} reduce
to the standard commutation relations of the Lie algebra
$\mathfrak{su}(2)$:
\begin{subequations}
  \begin{align}
    [\Tilde{J}_{+},\Tilde{J}_{-}]&=2\Tilde{J}_{z},\label{eq:3.7a}\\
    [\Tilde{J}_{z},\Tilde{J}_{\pm}]&=\pm\Tilde{J}_{\pm}.\label{eq:3.7b}
  \end{align}
\end{subequations}
The quadratic Casimir operator is therefore given by
\begin{subequations}
  \begin{align}
    C
    &=\Tilde{J}_{-}\Tilde{J}_{+}+\Tilde{J}_{z}(\Tilde{J}_{z}+1)=\frac{1}{\beta}J_{-}J_{+}+J_{z}^{2}-\frac{1}{4}=\frac{1}{\beta}H(J_{z})+J_{z}^{2}-\frac{1}{4}\label{eq:3.8a}\\
    &=\Tilde{J}_{+}\Tilde{J}_{-}+\Tilde{J}_{z}(\Tilde{J}_{z}-1)=\frac{1}{\beta}J_{+}J_{-}+(J_{z}-1)^{2}-\frac{1}{4}=\frac{1}{\beta}\Tilde{H}(J_{z})+(J_{z}-1)^{2}-\frac{1}{4},\label{eq:3.8b}
  \end{align}
\end{subequations}
which commutes with all the generators.

Now, let $|j,g\rangle$ be a simultaneous eigenstate of $C$ and $J_{z}$
that satisfies the eigenvalue equations
\begin{subequations}
  \begin{align}
    C|j,g\rangle&=\left(j^{2}-\frac{1}{4}\right)|j,g\rangle,\label{eq:3.9a}\\
    J_{z}|j,g\rangle&=g|j,g\rangle,\label{eq:3.9b}
  \end{align}
\end{subequations}
where $j\geq\tfrac{1}{2}$. We assume that the eigenstate $|j,g\rangle$
satisfies the normalization condition $\||j,g\rangle\|=1$, where the
norm $\||\ast\rangle\|=\sqrt{\langle\ast|\ast\rangle}$ is defined
through the inner product \eqref{eq:3.2}. Note that the commutation
relations \eqref{eq:3.5b} imply the following ladder equations:
\begin{align}
  J_{\pm}|j,g\rangle\propto|j,g\pm1\rangle.\label{eq:3.10}
\end{align}
The coefficients of proportionality can be determined by computing the
norms $\|J_{\pm}|j,g\rangle\|$. By using the relations
$J_{-}J_{+}=\beta(C-J_{z}^{2}+\frac{1}{4})$ and
$J_{+}J_{-}=\beta(C-(J_{z}-1)^{2}+\frac{1}{4})$ we find
\begin{subequations}
  \begin{align}
    \|J_{+}|j,g\rangle\|^{2}&=\beta(j^{2}-g^{2})\geq0,\label{eq:3.11a}\\
    \|J_{-}|j,g\rangle\|^{2}&=\beta(j^{2}-(g-1)^{2})\geq0,\label{eq:3.11b}
  \end{align}
\end{subequations}
where the inequalities follow from the positivity of the norms. These
equations not only fix the coefficients of proportionality but also
determine the possible values of $j$ and $g$. In fact, it is easy to
see that the constraints $j^{2}-g^{2}\geq0$ and $j^{2}-(g-1)^{2}\geq0$
together with the ladder equations \eqref{eq:3.10} are compatible with
each other if and only if $j$ is quantized as follows:
\begin{align}
  j\in\{\tfrac{1}{2},1,\tfrac{3}{2},2,\cdots\}.\label{eq:3.12}
\end{align}
Once such $j$ is given the eigenvalue $g$ takes the following values:
\begin{align}
  g\in\{j,j-1,j-2,\cdots,1-j\},\label{eq:3.13}
\end{align}
and the normalized eigenstate $|j,g\rangle$ is found to be of the form
\begin{align}
  |j,g\rangle=\sqrt{\frac{\Gamma(j+g)}{\beta^{j-g}\Gamma(2j)\Gamma(j-g+1)}}(J_{-})^{j-g}|j,j\rangle,\label{eq:3.14}
\end{align}
where $|j,j\rangle$ is the highest weight state that satisfies the
condition $J_{+}|j,j\rangle=0$.

\begin{figure}[t]
  \centering%
  \input{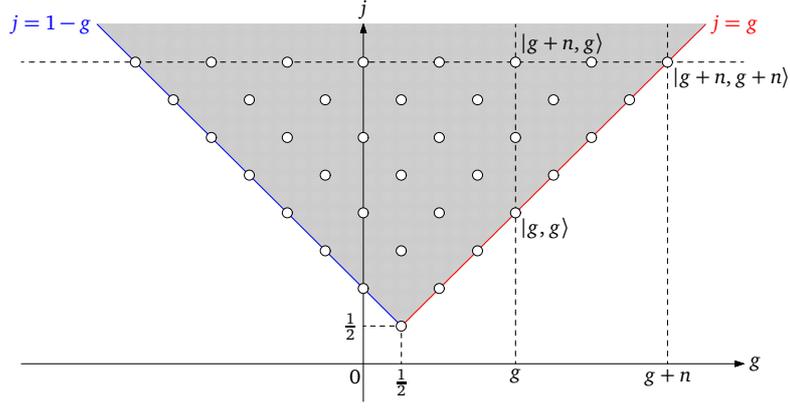}
  \caption{Unitary representations of the potential algebra
    $\mathfrak{su}(2)$ and the energy spectrum. White circles
    represent the states $\{|j,g\rangle\}$.}
  \label{figure:1}
\end{figure}

Now it is easy to solve the original spectral problem. To see this,
let us first note that, thanks to the relation
$H(g)=\beta(C-J_{z}^{2}+\frac{1}{4})$, the eigenvalue equation
\eqref{eq:3.9a} can be recast into the following form:
\begin{align}
  H(g)|j,g\rangle=\beta(j^{2}-g^{2})|j,g\rangle.\label{eq:3.15}
\end{align}
Now let $g\in\{\frac{1}{2},1,\frac{3}{2},2,\cdots\}$ be fixed. Then
the ground state of the Hamiltonian $H(g)$ corresponds to the state
$|g,g\rangle$ and the $n$th excited state of $H(g)$ corresponds to the
state $|g+n,g\rangle$; see figure \ref{figure:1}. The energy
eigenvalue $E_{n}$ of the $n$th excited state can therefore be read by
just substituting $j=g+n$ in \eqref{eq:3.15}. Thus we find
\begin{align}
  E_{n}=\beta\left[(g+n)^{2}-g^{2}\right]=(n^{2}+2ng)\beta,\quad n\in\{0,1,2,\cdots\},\label{eq:3.16}
\end{align}
which exactly coincides with \eqref{eq:2.9}.

The normalized energy eigenfunction can also be obtained from the
representation theory. To see this, let us first substitute $j=g+n$ in
\eqref{eq:3.14}:
\begin{align}
  |g+n,g\rangle=\sqrt{\frac{\Gamma(2g+n)}{\beta^{n}n!\Gamma(2g+2n)}}(J_{-})^{n}|g+n,g+n\rangle.\label{eq:3.17}
\end{align}
Let $\Psi_{g+n,g}(p,\theta)$ be a wavefunction corresponding to the
state $|g+n,g\rangle$. Then, it follows from \eqref{eq:3.1a} and
\eqref{eq:3.9b} that the $\theta$-dependence of
$\Psi_{g+n,g}(p,\theta)$ is just the plane wave
$\mathrm{e}^{ig\theta}$. Thus one may write
$\Psi_{g+n,g}(p,\theta)=\psi_{g+n,g}(p)\frac{\mathrm{e}^{ig\theta}}{\sqrt{2\pi}}$,
where $\psi_{g+n,g}(p)$ gives the normalized energy eigenfunction of
the eigenvalue $E_{n}$ of the Hamiltonian $H(g)$. Noting that
$\Psi_{g+n,g}(p,\theta)\propto(J_{-})^{n}\Psi_{g+n,g+n}(p,\theta)=\Bar{A}(J_{z})\mathrm{e}^{-i\theta}\Bar{A}(J_{z})\mathrm{e}^{-i\theta}\cdots\Bar{A}(J_{z})\mathrm{e}^{-i\theta}\psi_{g+n,g+n}(p)\frac{\mathrm{e}^{i(g+n)\theta}}{\sqrt{2\pi}}=\frac{\mathrm{e}^{ig\theta}}{\sqrt{2\pi}}\Bar{A}(g)\Bar{A}(g+1)\cdots\Bar{A}(g+n-1)\psi_{g+n,g+n}(p)$,
which follows from \eqref{eq:3.1c}, we find the following normalized
energy eigenfunction:\footnote{The eigenfunction \eqref{eq:3.18} can
  be written in terms of the Gegenbauer polynomial. Indeed, by
  introducing a new dimensionless variable
  $\xi=\frac{\sqrt{\beta}p}{1+\beta p^{2}}$ we find
  $\psi_{g+n,g}\propto(1-\xi^{2})^{-\frac{g-1}{2}}\frac{d^{n}}{d\xi^{n}}(1-\xi^{2})^{g+n-\frac{1}{2}}\propto(1-\xi^{2})^{\frac{g}{2}}C^{g}_{n}(\xi)$,
  where $C^{g}_{n}$ stands for the Gegenbauer polynomial of degree
  $n$.}
\begin{align}
  &\psi_{g+n,g}(p)\nonumber\\
  &=\sqrt{\frac{\Gamma(2g+n)}{\beta^{n}n!\Gamma(2g+2n)}}\Bar{A}(g)\Bar{A}(g+1)\cdots\Bar{A}(g+n-1)\psi_{g+n,g+n}(p)\nonumber\\
  &=(-1)^{n}\sqrt{\frac{\Gamma(2g+n)}{\beta^{n}n!\Gamma(2g+2n)}}(1+\beta p^{2})^{\frac{g-1}{2}}\left[(1+\beta p^{2})^{\frac{3}{2}}\frac{d}{dp}\right]^{n}(1+\beta p^{2})^{-\frac{g+n-1}{2}}\psi_{g+n,g+n}(p),\label{eq:3.18}
\end{align}
where in the second equality we have used \eqref{eq:2.5b}. Here
$\psi_{g+n,g+n}$ is the normalized solution to the first-order
differential equation $A(g+n)\psi_{g+n,g+n}(p)=0$, which turns out to
be given by
\begin{align}
  \psi_{g+n,g+n}(p)=\left(\frac{\beta}{\pi}\right)^{\frac{1}{4}}\sqrt{\frac{\Gamma(\frac{g+n+2}{2})}{\Gamma(\frac{g+n+1}{2})}}(1+\beta p^{2})^{-\frac{g+n}{2}}.\label{eq:3.19}
\end{align}

To summarize, we have solved the spectral problem only through the
unitary representations of the Lie algebra $\mathfrak{su}(2)$. The
defect of this approach, however, is that the representation theory
works only for $g\in\{\frac{1}{2},1,\frac{3}{2},2,\cdots\}$, though
the original eigenvalue equation \eqref{eq:2.8} can be solved for any
real $g$.

\section{Examples}
\label{section:4}
\paragraph{Example \#1.}
Having obtained the spectrum of the two-parameter family of
momentum-space Hamiltonian \eqref{eq:1.1}, we now proceed to examine
specific cases. The first example is that of the one-dimensional
harmonic oscillator within the minimal length uncertainty
formalism. The Hamiltonian of this system is given by
\begin{equation}
  H=\frac{1}{2m}p^{2}+\frac{1}{2}m\omega^{2}x^2.\label{eq:4.1}
\end{equation}
It follows from \eqref{eq:2.3} that this Hamiltonian can be expressed
in momentum space as
\begin{equation}
  H=\frac{m\hbar^{2}\omega^{2}}{2}\left[-(1+\beta p^{2})\frac{d}{dp}(1+\beta p^{2})\frac{d}{dp}+\frac{1}{m^{2}\hbar^{2}\omega^{2}}p^{2}\right].\label{eq:4.2}
\end{equation}
Now comparing \eqref{eq:4.2} with \eqref{eq:2.6a} we find
\begin{equation}
  g=\frac{1}{2}+\sqrt{\frac{1}{4}+\frac{1}{m^{2}\hbar^{2}\omega^{2}\beta^{2}}},\label{eq:4.3}
\end{equation}
where we have assumed $g$ is positive. The energy eigenvalue then
reads
\begin{align}
  E_{n}
  &=\frac{m\hbar^{2}\omega^{2}}{2}\left[(n^{2}+2ng)\beta+g\beta\right]\nonumber\\
  &=\hbar\omega\left(n+\frac{1}{2}\right)\left[\frac{m\hbar\omega\beta}{2}+\sqrt{1+\left(\frac{m\hbar\omega\beta}{2}\right)^{2}}\right]+\frac{1}{2}m\hbar^{2}\omega^{2}\beta n^{2},\quad n\in\{0,1,2,\cdots\},\label{eq:4.4}
\end{align}
which exactly coincides with the known result \cite{Kempf:1994su}. The
normalized energy eigenfunctions can be given by \eqref{eq:3.18} and
\eqref{eq:3.19} under the substitution \eqref{eq:4.3}. We note that,
due to the $n^{2}$ dependence, the deviation of the energy levels from
$\hbar\omega(n+\frac{1}{2})$ would become large for large $n$, from
which one could discuss the bound on the value of the minimal length
uncertainty $\hbar\sqrt{\beta}$; see, e.g., \cite{Chang:2001kn}.

\paragraph{Example \#2.}
Let us next consider a problem of relativistic quantum mechanics,
namely that of the minimum length Dirac oscillator
\cite{Nouicer:2006xua}. The eigenvalue equation for the Dirac
oscillator is given by \cite{Nouicer:2006xua}
\begin{equation}
  H\psi\equiv\left[c\sigma_{y}(p-i\sigma_{z}m\omega x)+\sigma_{z}mc^2\right]\psi=E\psi,\label{eq:4.5}
\end{equation}
where
$\sigma_{y}=\left(\begin{smallmatrix}0&-i\\i&0\end{smallmatrix}\right)$,
$\sigma_{z}=\left(\begin{smallmatrix}1&0\\0&1\end{smallmatrix}\right)$,
and $x$ and $p$ are given by \eqref{eq:2.3}. Writing
$\psi=\left(\begin{smallmatrix}f_{1}\\ f_{2}\end{smallmatrix}\right)$,
one finds that the eigenvalue equation for the upper component $f_{1}$
can be written as
\begin{equation}
  \left[-(1+\beta p^{2})\frac{d}{dp}(1+\beta p^{2})\frac{d}{dp}+\frac{1-m\hbar\omega\beta}{m^{2}\hbar^{2}\omega^{2}}p^{2}\right]f_{1}=\frac{1}{m^{2}\hbar^{2}\omega^{2}}\left(\frac{E^{2}-m^{2}c^{4}}{c^{2}}+m\hbar\omega\right)f_{1}.\label{eq:4.6}
\end{equation}
Now again comparing \eqref{eq:4.6} with \eqref{eq:2.6a} we find
\begin{equation}
  g=\frac{1}{m\hbar\omega\beta},\label{eq:4.7}
\end{equation}
where we have again assumed $g$ is positive. Setting
$\frac{1}{m\hbar^{2}\omega^{2}}(\frac{E_{n}^{2}-m^{2}c^{4}}{c^{2}}+m\hbar\omega)=(n^{2}+2gn)\beta+g\beta$
and solving this with respect to $E_{n}$ we get
\begin{align}
  E_{n}=\pm mc^{2}\sqrt{1+\frac{\hbar^{2}\omega^{2}\beta}{c^{2}}n^{2}+\frac{2\hbar\omega}{mc^{2}}n},\quad n\in\{0,1,2,\cdots\},\label{eq:4.8}
\end{align}
which is in perfect agreement with the known result
\cite{Nouicer:2006xua}.

It should be noted that we could have also considered the eigenvalue
equation for the lower component $f_{2}$, in which case the
Hamiltonian should be identified with \eqref{eq:2.6b}. Clearly the
potential algebra is the same as the Lie algebra $\mathfrak{su}(2)$
once the identification of the parameter is made. Note also that the
energy eigenfunctions are basically the same form as \eqref{eq:3.18}
and \eqref{eq:3.19} but the normalization constant should be
different, because in the Dirac oscillator the normalization condition
should be
$\int_{-\infty}^{\infty}\frac{dp}{1+\beta
  p^{2}}(|f_{1}|^{2}+|f_{2}|^{2})=1$.

\subsubsection*{Acknowledgement}
One of the authors (PR) would like to thank the Institute of Quantum
Science at Nihon University for supporting a visit during which part
of this work was done.

\appendix
\section{Two realizations of the potential algebra}
\label{appendix:A}
There have been proposed two approaches to the algebraic description
of shape invariance
\cite{Balantekin:1997mg,Gangopadhyaya:1998ccj}. However, these two
approaches are essentially equivalent. Focusing on our specific
example, in this appendix we shall discuss the equivalence between the
algebraic descriptions of shape invariance proposed by Balantekin
\cite{Balantekin:1997mg} and by Gangopadhyaya \textit{et al.}
\cite{Gangopadhyaya:1998ccj}. To this end, let us first introduce two
new operators $\Hat{\theta}$ and $\Hat{n}$ which are canonical
conjugate with each other and satisfy the following commutation
relations:
\begin{align}
  [\Hat{\theta},\Hat{n}]=i.\label{eq:A.1}
\end{align}
We then introduce the following operators:
\begin{subequations}
  \begin{align}
    J_{z}&=\Hat{n},\label{eq:A.2a}\\
    J_{+}&=\mathrm{e}^{i\Hat{\theta}}A(\Hat{n})=A(\Hat{n}-1)\mathrm{e}^{i\Hat{\theta}},\label{eq:A.2b}\\
    J_{-}&=\Bar{A}(\Hat{n})\mathrm{e}^{-i\Hat{\theta}}=\mathrm{e}^{-i\Hat{\theta}}\Bar{A}(\Hat{n}-1),\label{eq:A.2c}
  \end{align}
\end{subequations}
where $A$ and $\Bar{A}$ are given in \eqref{eq:2.5a} and
\eqref{eq:2.5b}. We note that the second equalities in \eqref{eq:A.2b}
and \eqref{eq:A.2c} follow from the identities
$\mathrm{e}^{\pm i\Hat{\theta}}\Hat{n}\mathrm{e}^{\mp
  i\Hat{\theta}}=\Hat{n}\mp 1$. A straightforward calculation then
gives
\begin{subequations}
  \begin{align}
    J_{-}J_{+}&=\Bar{A}(\Hat{n})A(\Hat{n})=-(1+\beta p^{2})\frac{d}{dp}(1+\beta p^{2})\frac{d}{dp}+\Hat{n}(\Hat{n}-1)\beta^{2}p^{2}-\Hat{n}\beta,\label{eq:A.3a}\\
    J_{+}J_{-}&=A(\Hat{n}-1)\Bar{A}(\Hat{n}-1)=-(1+\beta p^{2})\frac{d}{dp}(1+\beta p^{2})\frac{d}{dp}+\Hat{n}(\Hat{n}-1)\beta^{2}p^{2}+(\Hat{n}-1)\beta,\label{eq:A.3b}
  \end{align}
\end{subequations}
from which we find
\begin{align}
  [J_{+},J_{-}]=(\Hat{n}-1)\beta+\Hat{n}\beta=2\beta\left(\Hat{n}-\frac{1}{2}\right)=2\beta\left(J_{z}-\frac{1}{2}\right).\label{eq:A.4}
\end{align}
Similarly, we have
\begin{subequations}
  \begin{align}
    J_{z}J_{+}&=\Hat{n}\mathrm{e}^{i\Hat{\theta}}A(\Hat{n})=\mathrm{e}^{i\Hat{\theta}}A(\Hat{n})(\Hat{n}+1),\label{eq:A.5a}\\
    J_{+}J_{z}&=\mathrm{e}^{i\Hat{\theta}}A(\Hat{n})\Hat{n},\label{eq:A.5b}\\
    J_{z}J_{-}&=\Hat{n}\Bar{A}(\Hat{n})\mathrm{e}^{-i\Hat{\theta}}=\Bar{A}(\Hat{n})\mathrm{e}^{-i\Hat{\theta}}(\Hat{n}-1),\label{eq:A.5c}\\
    J_{-}J_{z}&=\Bar{A}(\Hat{n})\mathrm{e}^{-i\Hat{\theta}}\Hat{n}.\label{eq:A.5d}
  \end{align}
\end{subequations}
Thus we get the following commutation relations:
\begin{align}
  [J_{z},J_{\pm}]=\pm J_{\pm}.\label{eq:A.6}
\end{align}
By redefining the operators as $\Tilde{J}_{z}=J_{z}-\frac{1}{2}$ and
$\Tilde{J}_{\pm}=\frac{1}{\sqrt{\beta}}J_{\pm}$, one can easily find
that the set of operators
$\{\Tilde{J_{z}},\Tilde{J}_{+},\Tilde{J}_{-}\}$ satisfies the standard
commutation relations of the Lie algebra $\mathfrak{su}(2)$.

Now let us specialize to the following two realizations of the
operators $\Hat{n}$ and $\Hat{\theta}$:
\begin{subequations}
  \begin{alignat}{3}
    \text{(Case A)}&\quad\Hat{\theta}=i\frac{\partial}{\partial n}&\quad&\&\quad\Hat{n}=n,&\label{eq:A.7a}\\
    \text{(Case
      B)}&\quad\Hat{\theta}=\theta&\quad&\&\quad\Hat{n}=-i\frac{\partial}{\partial\theta}.&\label{eq:A.7b}
  \end{alignat}
\end{subequations}
It is obvious that these two realizations satisfy the commutation
relations $[\Hat{\theta},\Hat{n}]=i$. Correspondingly, we have the
following two realizations of the $SU(2)$ generators:
\begin{subequations}
  \begin{align}
    \text{(Case A)}
    &\quad
      \begin{cases}
        J_{z}=n,\\
        J_{+}=\mathrm{e}^{-\frac{\partial}{\partial n}}A(n),\\
        J_{-}=\Bar{A}(n)\mathrm{e}^{\frac{\partial}{\partial n}},\\
      \end{cases}\label{eq:A.8a}\\
    \text{(Case B)}
    &\quad
      \begin{cases}
        J_{z}=-i\frac{\partial}{\partial\theta},\\
        J_{+}=\mathrm{e}^{i\theta}A(-i\frac{\partial}{\partial\theta}),\\
        J_{-}=\Bar{A}(-i\frac{\partial}{\partial\theta})\mathrm{e}^{-i\theta}.\\
      \end{cases}\label{eq:A.8b}
  \end{align}
\end{subequations}
Note that eqs.~\eqref{eq:A.7a} and \eqref{eq:A.8a} correspond to the
operators discussed by Balantekin
\cite{Balantekin:1997mg}. Eqs.~\eqref{eq:A.7b} and \eqref{eq:A.8b}, on
the other hand, correspond to the operators discussed by Gangopadhyaya
\textit{et al.} \cite{Gangopadhyaya:1998ccj}.\footnote{Note that
  eqs.~\eqref{eq:A.7a}--\eqref{eq:A.8b} do not exactly coincide with
  the notations and prescriptions of \cite{Balantekin:1997mg} and
  \cite{Gangopadhyaya:1998ccj}. But the essential ideas are the same
  as those proposed in \cite{Balantekin:1997mg} and
  \cite{Gangopadhyaya:1998ccj}.} Now it is obvious that these two
descriptions are merely the choice of the realizations for the
operators $\Hat{n}$ and $\Hat{\theta}$ and hence essentially
equivalent.

\bibliographystyle{utphys}%
\bibliography{bibliography}%
\end{document}